# Manybody *GW* calculation of the oxygen vacancy in ZnO

Stephan Lany and Alex Zunger

*National Renewable Energy Laboratory, Golden, CO 80401*

(dated: 2/8/2010)

Density functional theory (DFT) calculations of defect levels in semiconductors based on approximate functionals are subject to considerable uncertainties, in particular due to inaccurate band gap energies. Testing previous correction methods by many-body *GW* calculations for the O vacancy in ZnO, we find that: (i) The *GW* quasi-particle shifts of the $V_\text{O}$ defect states increase the spitting between occupied and unoccupied states due to self-interaction correction, and do not reflect the conduction versus valence band character. (ii) The *GW* quasi-particle energies of charged defect states require important corrections for supercell finite size effects. (iii) The *GW* results are robust with respect to the choice of the underlying DFT or hybrid-DFT functional, and the (2+/0) donor transition lies below mid-gap, close to our previous prediction employing rigid band edge shifts.

Density functional theory (DFT) in the local density or gradient corrected approximations (LDA or GGA) has long formed the basis for most theoretical accounts of defects in semiconductors and insulators. However, the underlying approximations for the electronic interactions lead to significant ambiguities, such as the ill-determined position of charge transition levels [1, 2] due to the notorious "band gap problem". These well-known limitations spurred the development of methods for correcting DFT results, ranging from "post-processor" corrections that are applied *after* DFT energies were calculated [1, 2, 3, 4] to empirical DFT corrections that are applied self-consistently [5, 6], and post-DFT methods like self-interaction correction [7] and hybrid-DFT [8, 11]. Alternatively, many-body perturbation theory based on the *GW* approximation [9] for the electron self-energy has been very successful for the prediction of quasi-particle energy spectra, i.e., the band-structures, of defect-free semiconductors and insulators [10]. It is expected that *GW* will set the benchmark also for defects [11, 12]. We choose here the classic case of the oxygen vacancy in ZnO as a system that has received a great deal of interest and debate in the literature [2, 3, 4, 5, 7, 8, 13]. The purpose of this work is to reevaluate DFT correction methods in view of *GW* quasi-particle energy calculations for the defect states of $V_O$. The charge-neutral vacancy $V_O^0$ introduces a doubly occupied $a_1^2$ level inside the band gap, and successive ionization leads to the $V_O^{1+}$ ($a_1^1$) and $V_O^{2+}$ ($a_1^0$) charge states, as illustrated in Fig. 1. The quantities of interest (and debate) are: (*i*) The donor levels which determine the electrical activity of $V_O$, i.e., the electrical transition energies ε(2+/1+) and ε(1+/0). Since, $V_O$ is a negative-*U* center [2, 3, 4, 5, 7, 8, 13], the most important transition energy is the position of the ε(2+/0) equilibrium transition level in the corrected band gap. (*ii*) The absolute formation energy of $V_O$ which determines the abundance of this defect in real ZnO materials. Reviews of the experimental literature on O vacancies are found in Refs. [3, 4].

*Obtaining structurally-relaxed transition energies from GW*: The *GW* method has been used mainly to determine quasi-particle (QP) energies within many-body perturbation theory, but not for calculation of total energies and structural relaxation (although first steps into this direction have been taken [14, 15]). Indeed, we do not attempt to calculate here the transition energies ε(*q/q'*) directly from *GW* total energies. As illustrated in Fig. 1, we determine instead separately the vertical (Franck-Condon) ionization energies $ε_O(q→q')$ and the subsequent structural relaxation energies $E_{rel}(q')$ in the final state.

While we have determined the vertical transition energies of $V_O$ in ZnO before in a DFT study [3], we now calculate the QP-energies of the defect states in *GW* to determine more accurately the vertical transitions relative to the band edges. The structural relaxation energies $E_{rel}$ are calculated within the underlying DFT or hybrid-DFT Hamiltonian. Note that the *GW* calculated QP-energies are used here to determine the electron removal energies, for which excitonic electron-hole interactions should not be included. Instead, the conduction band minimum (CBM) serves as a *distant* reservoir for free electrons ("*e*" in Fig. 1d) and defines an energy reference for the defect QP energies.

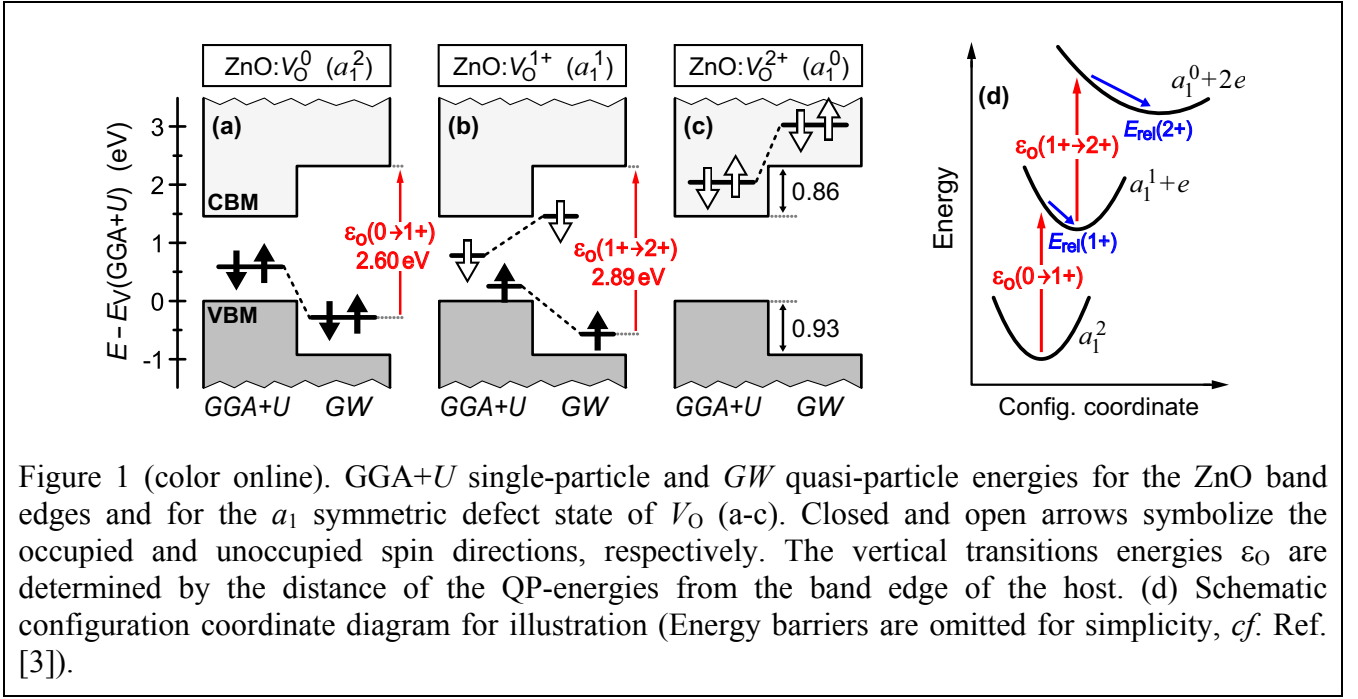

Figure 1 (color online). GGA+*U* single-particle and *GW* quasi-particle energies for the ZnO band edges and for the $a_1$ symmetric defect state of $V_O$ (a-c). Closed and open arrows symbolize the occupied and unoccupied spin directions, respectively. The vertical transitions energies $\varepsilon_O$ are determined by the distance of the QP-energies from the band edge of the host. (d) Schematic configuration coordinate diagram for illustration (Energy barriers are omitted for simplicity, *cf*. Ref. [3]).

The equilibrium transition level $\varepsilon(q/q')$ are determined from the sum of the vertical transition energies $\varepsilon_O$ and the relaxation energies $E_{rel}$. In principle, one can either use the electron *removal energies* $\varepsilon_O(q \rightarrow q+1)$ from occupied states to determine the successive transition energies towards *higher* charge states [e.g., $V_O^0(a_1^2) \rightarrow V_O^{1+}(a_1^1) \rightarrow V_O^{2+}(a_1^0)$, see Fig. 1d], or, as done in Ref. [12], one can use the electron *addition energies* $\varepsilon_O(q \rightarrow q-1)$ into unoccupied states to determine the successive transition energies towards *lower* charge states [e.g., $V_O^{2+}(a_1^0) \rightarrow V_O^{1+}(a_1^1) \rightarrow V_O^0(a_1^2)$]. Both ways should lead to the same result for the $\varepsilon(q/q')$. We here choose the former option, because in ZnO, the $a_1$ defect state of $V_O^{2+}$ occurs as a broad resonance deep inside the conduction band (see Fig. 1c), impairing the accurate determination of the electron addition energy. In addition, since the QP energies of charged defects in supercells are subject to electrostatic finite-size effects (see below), it is desirable to avoid higher charge states.

*Methods.* The present calculations are performed in the projector augmented wave (PAW) framework of the VASP code [16] which includes recent implementations of hybrid-DFT [17] and *GW* [18]. Supercell finite-size effects are treated as described in Refs. [2, 19]. For the sake of computational feasibility we consider here the metastable zinc-blende (ZB) phase of ZnO [20], which has a higher symmetry but otherwise has very similar properties as wurtzite (WZ) ZnO [21]. For the underlying DFT calculation, needed to determine the wavefunctions for the subsequent *GW* calculation and for the relaxation energies $E_{\rm rel}$, we use the GGA parameterization of Ref. [22], and employ the DFT+*U* method [23] for the Zn-*d* electrons with $U-J = 6$ eV, as in previous GGA+*U* calculations of defects in ZnO [6]. We refer to *GW* based on GGA+*U* as "*GW*-GGA+*U*". The motivation for the choice of the GGA+*U* method is that, as shown in Fig. 1, the single-particle defect energies relative to the band edges are described qualitatively correctly for the all three charge states of $V_{\rm O}$. In contrast, in GGA (without U) the $V_{\rm O}$ defect state in the 1+ state exhibits a spurious hybridization with the conduction band, which leads to an erroneous charge and spin density and to incorrect atomic relaxation [2] and precludes the calculation of *GW* quasi-particle energies based on GGA wavefunctions [24]. For comparison, we perform the same type of *GW* calculations also based on the HSE hybrid-DFT functional [25] ("*GW*-HSE"), using $\alpha = 0.25$ for the fraction of Fock exchange and $\mu = 0.2$ Å$^{-1}$ for the range separation parameter.

For computational economy, we employ a relatively soft PAW pseudopotential (PP) for oxygen (PAW radius: $R = 1.0$ Å), which has been tested for ZnO before in DFT [2] and hybrid-DFT [8] calculations (the error in the binding energy of the O$_2$ molecule due to the soft PP [2, 8] has been corrected). The two atomic ZB cell of defect-free ZnO was calculated using a Γ centered 8×8×8 k-mesh and a total of 144 bands. The *GW* QP energies in the 64 atom supercells are calculated with a Γ centered 2×2×2 k-mesh and a total of 2048 bands, and the response functions are determined only at the Γ point. We tested the effect of these reductions of computational parameters in smaller cells of pure ZnO, and expect that the resulting uncertainty should not exceed 0.2 eV for the QP energies of the $V_{\rm O}$ defect states relative to the band edges.

*Finite-size correction for quasi-particle energies of defects.* We have previously addressed [2, 19] image charge corrections for *total energies* of charged supercells. In general, however, also the DFT *single-particle* energies or the *GW quasi-particle* energies require corrections, if the defect state is localized. In this case, the energy $e_D$ of the defect state is shifted by an amount $\Delta e_D = \Delta V_D(\mathbf{R}_0)$ due to the electrostatic potential $\Delta V_D$ that is created by the charged defect images and by the compensating background at the site $\mathbf{R}_0$ of the defect. In order to *illustrate* the importance of this type of finite-size effects, we show in Fig. 2 the GGA+*U* calculated single-particle energy $e_D$ of the unoccupied $a_1$ state of $V_O^{2+}$ as a function of the supercell size. For demonstration purposes, the atomic configuration is here constrained such the $a_1$ state lies within the GGA+*U* band gap within the series of supercells between 64 and 1000 atoms (in a fully relaxed calculation, the $a_1$ state of $V_O^{2+}$ lies above the CBM, see Fig. 1c). Potential alignment effects [2, 19] have been taken into account to determine the energy $e_D$ relative to the valence band maximum (VBM) of the defect-free ZnO host. We now determine a finite-size correction for $e_D$ by calculating the potential $\Delta V_D$ at the site $\mathbf{R}_0$ of the O vacancy. Here, the charged vacancy images in neighboring supercells are approximated as point charges (the self-potential due to the charged vacancy at $\mathbf{R}_0$ is excluded). The dielectric screening is taken into account by dividing the bare Coulomb potential by the dielectric constant. As seen in Fig. 2 from the GGA+*U* calculations up to 1000 atom supercells, this correction accurately removes the finite size effects. The *GW* calculated QP energies of the defect-state will be affected in exactly the same manner, since this size-dependence of the GGA+*U* single particle energies of the localized defect state results from purely electrostatic interaction between supercells (not from interactions due to overlapping defect-state wave-functions). Therefore, we apply analogous corrections to the *GW* QP energies of $V_O^{1+}$ and $V_O^{2+}$ shown in Figs. 1b and c. Note, however, that the QP energy of the doubly charged vacancy $V_O^{2+}$ is not needed for the prediction for the ε(2+/0) equilibrium transition (*cf.* Fig. 1)

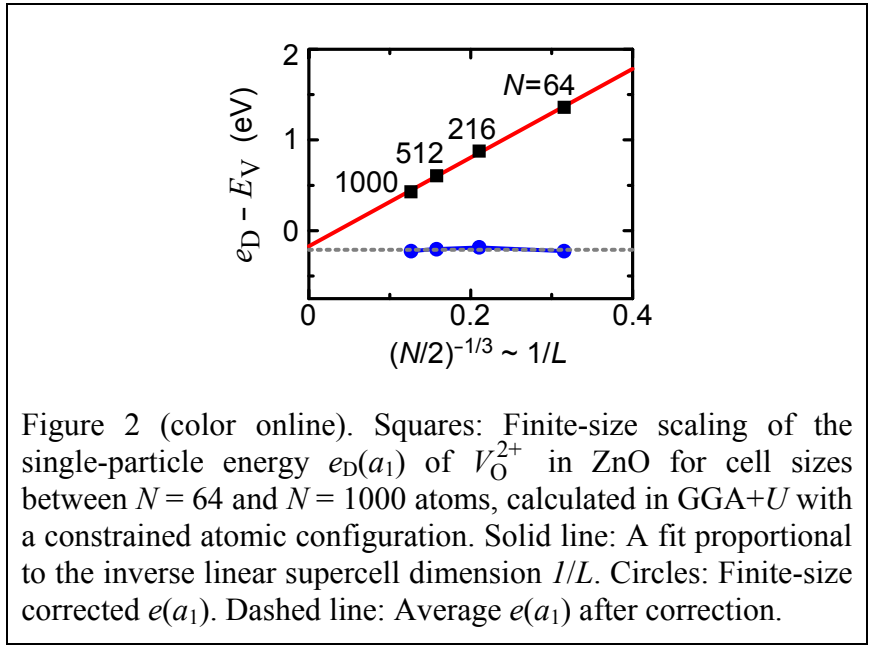

Figure 2 (color online). Squares: Finite-size scaling of the single-particle energy $e_D(a_1)$ of $V_O^{2+}$ in ZnO for cell sizes between $N = 64$ and $N = 1000$ atoms, calculated in GGA+$U$ with a constrained atomic configuration. Solid line: A fit proportional to the inverse linear supercell dimension $1/L$. Circles: Finite-size corrected $e(a_1)$. Dashed line: Average $e(a_1)$ after correction.

*Quasi-particle energies of the $V_O$ defect states.* The $GW$ QP-energies $e^{GW}$ relative to the initial DFT (or hybrid-DFT) eigen-energies $e^{DFT}$ values are determined as

$$e_n^{GW} = e_n^{DFT} + \text{Re}\langle \psi_n^{DFT} | \Sigma(e_n^{GW}) - V_{xc}^{DFT} | \psi_n^{DFT} \rangle, \tag{1}$$

where the initial DFT wavefunctions $\psi_n^{DFT}$ ($n$ = band index) are kept constant, and the $GW$ self-energy $\Sigma$ is determined by iteratively updating (4 times) the eigenvalues in $G$ and in $W$ (both in bulk and defect calculations). The resulting $GW$ band gap of ZB ZnO is 3.25 eV ($GW$-GGA+$U$) and 3.34 eV ($GW$-HSE), in agreement with previous $GW$ calculations [14, 26]. Figure 1 shows the QP energy shifts for both the ZnO band edges and for the $a_1$ symmetric $V_O$ defect level, whose energy strongly depends on the $V_O$ charge state and the respective atomic configuration of $V_O$ [3]. Remarkably, the $a_1$ state tracks the shift of the VBM ($E_V$) if it is occupied, but it tracks the CBM ($E_C$) if it is unoccupied. This $GW$ result is at variance with the expectation that the defect levels would shift in proportion to their CBM vs. VBM wavefunction characters [4]. Instead, the QP energy shifts appear to reflect the self-interaction correction which increases the splitting between occupied and unoccupied states, even when the wavefunction character is similar (see Fig. 1b). This finding lends some justification to the band gap correction via a rigid shift of the conduction band [1]: To the extend that the occupied defect QP energies track the valence band, the vertical electron removal energies remain invariant relative to the VBM. Since the cation-*d* states generally experience a larger self-interaction error than, e.g., the anion-*p* states, it is often practical to use DFT+U for cation-*d* states before shifting the conduction band [3, 13].

In case of $V_O^{2+}$, the $a_1$ state forms a broad resonance deep inside the conduction band, both in GGA+*U* and *GW*, which confirms the expectation [3, 13] that $V_O$ would be a source of persistent photoconductivity. (Shown in Fig. 1c is the energy corresponding to the center-of-mass of the vacancy-site projected *s*-like density of states in a fully relaxed supercell, as described in Ref. [3] for the case of a standard LDA calculation.)

*GW corrected thermodynamic transition energies.* We now turn to combine the vertical transition energies $\varepsilon_O$ from *GW* and the relaxation energies $E_{rel}$ from GGA+*U* to determine the equilibrium transition energies $\varepsilon(q/q')$. Vertical transitions require, however, special care in correcting finite-size effects, due to the simultaneous presence of electronic and ionic screening [19]: Consider, for example, the optical transition $V_O^{1+} \rightarrow V_O^{2+} + e$ (see Fig. 1b), where the atomic configuration of the final $V_O^{2+}$ state is constrained to that of the initial 1+ state. Here, the electronic screening attenuates the 2+ defect charge, but the ionic contribution still reflects the screening of the initial 1+ state, which makes it difficult to correct the image charge interaction of such intermediate states (this problem does not exist if the initial state is charge-neutral). In order to avoid these ambiguities, we apply the following a two-step procedure: First we calculate the $V_O$ transition levels from the *GW* QP energies and the (hybrid-) DFT relaxation energies by constraining the structural relaxation to the first two atomic shells around $V_O$ (Fig. 3a). This eliminates ionic screening of the interaction between supercells, and we can use the calculated *electronic* dielectric constant to determine the image charge corrections. The vertical $\varepsilon_O(0 \rightarrow 1+)$ and $\varepsilon_O(1+ \rightarrow 2+)$ energies under this constraint are, respectively, 2.60 and 2.89 eV in *GW*-GGA+*U* (see Fig. 1b and d), or 2.88 and 3.43 eV in *GW*-HSE.

In a second step (Fig. 3b), we then calculate the (hybrid-) DFT supercell energies for all charge states without any constraint, where we can use the total dielectric constant due to combined *electronic and ionic* screening [19], where we use the experimental value of 8.1. The removal of the constraint lowers the formation energy mostly for the 2+ state and leads to a negative-*U* behavior (as before, see Refs. [3, 4, 7, 8]) with a $\varepsilon(2+/0)$ transition at $E_V+1.36$ eV (*GW*-GGA+*U*), as shown in Fig. 3b. In *GW*-HSE, we obtain the transition at $E_V+1.66$ eV, noting that the difference mainly reflects the larger lattice constant of HSE (which is close to that of GGA) compared to GGA+$U$ [27]. Thus, apart from the effect of the lattice parameter, the *GW* results based on GGA+$U$ and HSE (see Table I) are essentially identical, and agree well with our previous DFT-corrected prediction at $E_V$+1.30 eV [13]. For the HSE hybrid functional ($\alpha$=0.25), we observed that the position of the $\varepsilon$(2+/0) level remains virtually constant in *GW*-HSE when measured relative to the VBM (see Table I). Since, however, *GW* shifts up the HSE calculated CBM by about 1 eV, the ionization energy for release of free electrons increases considerably due to the *GW* corrections. On the other hand, hybrid-DFT calculations using the HSE hybrid functional with an increased fraction of Fock exchange so as to reproduce the experimental band gap of ZnO [8] gave a transition level higher in the gap at $E_V$+2.2 eV (we obtain here $E_V$+2.34 eV for HSE with $\alpha$ = 0.40). The present *GW* results suggest, however, that the $V_O$ defect level is better described by using the standard form of HSE ($\alpha$=0.25) plus a rigid shift of about 1 eV for the CBM.

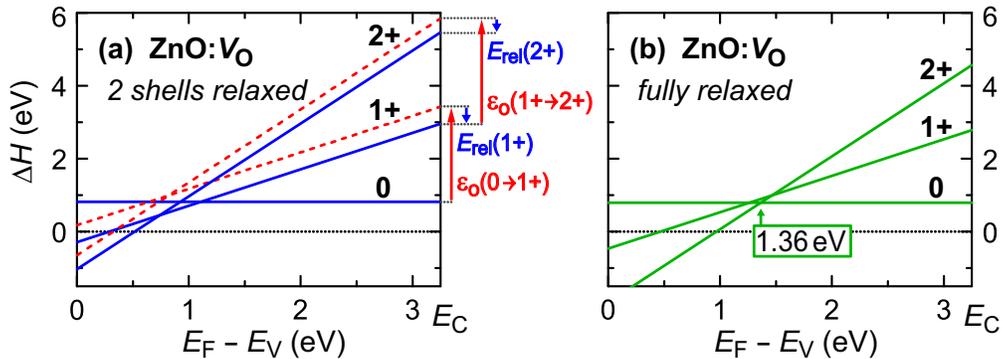

Figure 3 (color online). Formation energy $\Delta H(E_F)$ of $V_O$ in ZnO under O-poor conditions, calculated using vertical (optical) transition energies $\varepsilon_O$ from *GW* and relaxation energies $E_{rel}$ from GGA+$U$. (a) Atomic relaxation is restricted to the first two nearest neighbor shells. (b) Formation energies obtained when full relaxation of all atoms is allowed.

*Absolute formation energies.* Besides the position of the defect level in the gap, previous calculations also differed about the absolute formation energy of $V_O$ [4, 5, 7, 8, 13], which, as mentioned above, is presently not accessible in the *GW* method. Here, we use the energy of the charge-neutral $V_O^0$ state, as calculated by GGA+$U$ or HSE, along with the transition levels $\varepsilon(q/q')$ as determined above, to obtain the absolute formation energies $\Delta H(V_O^q)$, as shown in Fig. 3b. In case of GGA+$U$, however, there exists an ambiguity, because the value of $U$ used for the ZnO compound is not suitable also for the elemental metallic phase of Zn. Therefore, we use for this case the optimized elemental reference energies of Ref. [28], which are determined so as to optimize the degree of error cancellation between the energies of the compound and that of the elemental constituents. Thus, in the Zn-rich/O-poor limit, we obtain $\Delta H(V_O^0) = 0.81$ eV based on GGA+$U$, which is close to the prediction of HSE irrespective of the value of the parameter α (Table I and Ref. [8]). Note, however, that by using the elemental reference energies of Ref. [28] for GGA+$U$ we better reconcile the experimental heat of formation of ZnO ($\Delta H_f = -3.63$ eV) than GGA [28] and hybrid-DFT (Ref. [8] and Table I), thereby describing better the $\Delta H(V_O)$ difference between the Zn-rich/O-poor and the Zn-poor/O-rich conditions.

Table I. Properties of zinc-blende ZnO in GGA+$U$, HSE, and in $GW$: The band gap $E_g$, the heat of formation $\Delta H_f$ of ZnO, the formation energy of $V_O$ under O-poor/Zn-rich conditions, and the thermodynamic ε(2+/0) transition level of $V_O$. All numbers in eV.

|  | $E_g$ | $\Delta H_f$ | $\Delta H(V_O^0)$ | ε(2+/0) |
|---|---|---|---|---|
| GGA+$U$ | 1.46 | $-3.74^a$ | $0.81^a$ | $E_V$+0.98 |
| $GW$-GGA+$U$ | 3.25 | - | - | $E_V$+1.36 |
| HSE | 2.34 | $-3.07$ | 0.96 | $E_V$+1.67 |
| $GW$-HSE | 3.34 | - | - | $E_V$+1.66 |

$^a$ elemental reference energies for GGA+$U$ are taken from Ref. [28]

*Conclusions.* We calculated the quasi-particle energies for the defect states of the O vacancy in ZnO within the $GW$ approximation based on DFT and hybrid DFT wavefunctions, paying particular attention to finite size effects for the QP energies of the charged defect states. The resulting thermodynamic ε(2+/0) donor transition lies consistently at or below mid-gap, irrespective of the underlying functional (GGA+$U$ or HSE), and agrees quite well with previous results [13], where the band gap error was corrected through rigid shifts of the band edges. The defect level of the O vacancy in ZnO predicted by the HSE hybrid-DFT functional is well described when using the standard parameter α=0.25 for the Fock exchange, plus a rigid shift (~1 eV) of the CBM. In contrast, adjusting the hybrid DFT parameters so to match the experimental band gap tends to move the $V_O$ defect level too close to the CBM.

**Acknowledgments**

This work is supported by the U.S. Department of Energy, Office of Science, Office of Basic Energy Sciences under Contract No. DE-AC36-08GO28308 to NREL. The Center of Inverse Design is a DOE Energy Frontier Research Center. The use of MPP capabilities at the National Energy Research Scientific Computing Center is gratefully acknowledged. S.L. thanks Georg Kresse for providing pseudopotentials for test purposes that were specifically created for *GW*-type calculations.